\documentclass[preprint2]{aastex}

\newcommand{\Note}[1]{
  \advance\textwidth-72pt\par
  \parbox{\textwidth}{
    \hfill\Putline
    {\small\sf#1\par}
    \Putline\par
    }
  \advance\textwidth+72pt\par
}
\newcommand{\Putline}{\rule{\the\textwidth}{1pt}  }
\begin{document}

\title{High Resolution X-Ray Spectra of Capella: Initial Results from the
Chandra High Energy Transmission Grating Spectrometer}

\author{C.~R.~Canizares, D.~P.~Huenemoerder, D.~S.~Davis, D.~Dewey, K.~A.~Flanagan, 
J.~Houck, T.~H.~Markert, H.~L.~Marshall, M.~L.~Schattenburg, N.~S.~Schulz,
M. Wise } \affil{MIT Center for Space Research\\
70 Vassar St., \\
Cambridge, MA  02139}
\author{J.~J.~Drake, N.~S.~Brickhouse}
\affil{Harvard-Smithsonian Center for Astrophysics\\
  60 Garden St.\\
  Cambridge, MA  02138}

\begin{abstract}
  High resolution spectra of the active binary Capella (G8 III + G1
  III) covering the energy range 0.4-8.0 keV (1.5-30\AA) show a large
  number of emission lines, demonstrating the performance of the
  HETGS.  A preliminary application of plasma diagnostics provides
  information on coronal temperatures and densities.  Lines arising
  from different elements in a range of ionization states indicate
  that Capella has plasma with a broad range of temperatures, from
  $\log T = 6.3~{\mathrm{to}}~ 7.2$, generally consistent with recent
  results from observations with the Extreme Ultraviolet Explorer
  (EUVE) and the Advanced Satellite for Cosmology and Astrophysics
  (ASCA).  The electron density is determined from He-like O~{\sc vii}
  lines, giving the value $N_e \sim 10^{10} \mathrm{cm^{-3}}$ at $T_e
  \sim 2 \times 10^6 \mathrm{K}$; He-like lines formed at higher
  temperatures give only upper limits to the electron density. The
  density and emission measure from O~{\sc vii} lines together
  indicate that the coronal loops are significantly smaller than the
  stellar radius.
\end{abstract}



\section{Introduction}

The active binary Capella ($\alpha$ Aurigae, HD~34029, HR~1708) was
observed with the High Energy Transmission Grating Spectrometer
(HETGS) on the Chandra X-Ray Observatory (CXO).  We present a first
analysis of the spectra with the goals of demonstrating the HETGS
performance, and of applying plasma diagnostics to infer physical
parameters of the Capella corona.  A complementary analysis of the
corona of Capella based on high resolution spectra obtained using the
CXO Low Energy Transmission Grating Spectrometer (LETGS) has been presented by
\citet{Brinkman2000}.
Further analysis of diagnostic emission lines from these and other
Chandra grating data of Capella are underway with the goal of
obtaining refined temperature-dependent emission measures, abundances,
and densities, leading to a self-consistent determination of the
coronal structure.

\paragraph{The Chandra HETGS:}
The High Energy Transmission Grating assembly \citep{Markert94, Canizares00}
consists of an array of periodic gold microstructures that can be
interposed in the converging X-ray beam just behind the Chandra High
Resolution Mirror Assembly.  When in place, the gratings disperse the
X-rays according to wavelength, creating spectra that are recorded
at the focal plane by the linear array of CCDs designated ACIS-S.
There are two different grating types, designated MEG and HEG,
optimized for medium and high energies (partially overlapping in
spectral coverage).  The HETGS provides spectral resolving power of
$\lambda/\Delta\lambda =100$-1000 for point sources (corresponding to
a line FWHM of about 0.02 \AA\ for MEG, and 0.01 \AA\ for HEG) and
effective areas of ~1-180 $\mathrm{cm^2}$ over the wavelength range
1.2-30~\AA\ (0.4-10 keV).  Multiple overlapping orders are separated
using the moderate energy resolution of the ACIS detector.  The HETGS
complements the LETGS, which is optimized for lower energy X-rays.
(For detailed descriptions of the instruments see {\tt
  http://chandra.harvard.edu}).

Preliminary analysis of in-flight calibration data including those
presented here indicates that the HETGS is performing as predicted
prior to the Chandra launch. The spectral resolution is as expected
and effective areas are within 10\% of the expected values except from
6--12\AA\ where there are systematic uncertainties of up to 20\%.
Ongoing calibration efforts will reduce these uncertainties.

\paragraph{The Coronal Structure of Capella:}

Capella is an active binary system comprised of  G1 and  G8 
giants in a 104 d orbit at a distance of 12.9 pc. The G1 star
rotates with an $\sim 8$ d  period  
\citep{Hummel94}.
Capella has been studied by many previous
X-ray telescopes, including Einstein \citep{Holt79, Swank81a,
Mewe82, Vedder83, Schmitt90},
EXOSAT \citep{Lemen89}; ROSAT
\citep{Dempsey93b}, Beppo-SAX \citep{Favata97}, and ASCA
\citep{BRICK00}. The fundamental parameters of Capella, some activity
indicators, and primary references may be found in \citet{CABS93}.

The corona of Capella appears intermediate in temperature, being
cooler than those of RS~CVn stars such as HR~1099 or II~Peg, but
significantly hotter than a less active star like Procyon. X-ray
observations obtained at low to moderate spectral resolution are
generally consistent with emission from an optically thin,
collisionally dominated plasma with two temperature components
\citep{Swank81a, Schmitt90}.  Spectra obtained by the Extreme
Ultraviolet Explorer (EUVE) have provided more discriminating
temperature diagnostics, showing plasma over a continuous range of
temperatures, with the peak emission measure near $\log T=6.8$
\citep{Dupree93, Schrijver95, BRICK00}.  
Simultaneous measurements using EUVE and ASCA spectra did not
require emission from plasma hotter than $log T = 7.3$ \citep{BRICK00}.
EUVE observations show variability by factors of 3 to 4 in lines 
formed above $log T \sim 7.0$ \citep{BRICK00, Dupree00}.

\citet{Dupree93} have estimated plasma electron densities in the range
from $4 \times 10^{11}$ to $10^{13} \mathrm{cm^{-3}}$ from lines of
Fe~{\sc xxi} formed near $10^{6.8} \mathrm{K}$, implying that the
scale of the emitting volume is $\sim 10^{-3} R_*$, although
\citet{GRIFF98} question the reliability of this diagnostic.
\citet{BRICK00} use EUV lines of Fe~{\sc xviii} to
constrain the optical depth in
the strong X-ray emission line, Fe~{\sc xvii} $\lambda$15.014, 
to $\tau < 3.6$.

From high-resolution UV spectra from the Hubble Space Telescope,
 \citet{Linsky98} concluded that both stars have
comparable coronal emission, 
based on measurements of the Fe~{\sc xvii} (1354\AA) coronal forbidden
line,
and that the plasma is magnetically confined.  Thus the ``corona'' of
Capella is actually a composite of two ``coronae.''

\section{Observations and Data Processing}

We combined data from three HETGS observations 
(from 1999 August 28, September 24 \& 25) for a total exposure of
89~ks.  Data were processed with the standard Chandra X-Ray Center
software (versions from July 29 ({\tt R4CU3UPD2}) and December 13
({\tt CIAO 1.1})).  The image of the dispersed spectrum is shown in
Figure~\ref{fig:image}.  
%
%
Each photon is assigned a dispersion angle, $\theta$, relative to the
undiffracted zero-order image. The angle is related to the
order, $m$, and wavelength, $\lambda$, through the grating mean
period, $P$, by the grating equation, $m\lambda=P\sin\theta$.
The spectral order is determined using the ACIS-S CCD pulse
height for each photon event (with wide latitude to avoid sensitivity
to variations in CCD gain or pulse height resolution).
The positive and negative first orders were summed separately for HEG and MEG
 for all observations and divided by the effective areas
to provide flux-calibrated spectra (Figure~\ref{fig:spectrum}).
%
%
listed in Table~\ref{tab:linelist}.  
%
%
the Fe~{\sc xvii} $\lambda15.01$ line strength is, within the uncertainties,
identical to that observed in 1979 with the Einstein crystal spectrometer by
\citet{Vedder83}, while the O {\sc viii} $L\alpha$ line is roughly half
the previous value.

\section{Coronal Diagnostics}
\paragraph{Emission Measure Distribution:}
Some properties of the coronal temperature structure can be deduced
from a preliminary analysis of the spectrum. The data warrant a full
analysis of the volume emission measure distribution with
temperature, $VEM(T)$, 
($VEM(T) ~\propto ~N_e^2~\times~V$
in which $N_e$ is the electron density of plasma at temperature $T$ which
occupies the volume, $V$), which will be the subject of a future paper. 

As 
Table~\ref{tab:linelist} illustrates, the spectrum contains lines from
different elements in a range of ionization states, demonstrating that the
emitting plasma has a broad range of
temperature.  Further evidence of multi-temperature emission comes
from two line ratios.  First, ratios of H-like to He-like
resonance lines, O~{\sc viii/vii}, Mg~{\sc xii/xi}, and Si~{\sc
xiv/xiii} indicate ionization ratios corresponding to $\log T$ = 6.55-6.60,
 6.75-6.85, and 6.95-7.00, respectively.  
Second, the
He-like ions provide  temperature-sensitive ratios
involving the resonance ($r$), forbidden ($f$) and
intersystem ($i$) lines
\citep{Gabriel69, Gabriel72, Pradhan81, Smith98}.
For the observed O~{\sc vii}, Mg~{\sc xi}, and Si~{\sc xiii} lines, the ratio
$(i+f)/r$ corresponds to temperatures $\log T=6.2-6.4$, $6.9-7.1$, and
$6.85-6.95$, respectively, using the theoretical models of Smith et
al. (1998, in the low density limit). In both cases, the ratios indicate
that the corona has a broad range of temperature.

An approximate upper envelope to the true $VEM$ distribution is given
by the family of curves formed by plotting the ratio of line strength
to corresponding emissivity for a collection of lines.  For a given
element, its abundance affects only the overall normalization of the
envelope of all lines from that element. 
For this initial analysis, we assumed Solar abundances
\citep{Anders89}, which is consistent with previous analyses
except possibly for Ne \citep{BRICK00}.

The VEM envelope of Figure~\ref{fig:vemt},
%
%
indicatates that plasma must be present over
nearly a decade in temperature.  The absence of lines from He-like and
H-like ions of Fe provides an upper limit to the $VEM$ above $\log
T=7.2$.  
Although the envelope does not trace closely the peaked 
distribution
derived from EUV lines, such a distribution is not excluded,

\paragraph{Density Diagnostics:}
The He-like $f/i$ ratio is primarily sensitive to density
Using the theoretical line ratios of 
Smith et al. (1998),
our measured O~{\sc vii} ratio of $2.9\pm0.4$
implies an electron density within the range 
$0.8$--$2\times10^{10}\,\mathrm{cm^{-3}}$. 
Similarly, the Mg~{\sc xi} and Si~{\sc xiii} ratios of $3.0\pm0.3$ and
$2.6\pm0.2$ give upper limits near $7\times10^{11}$ and
$1\times10^{12}\,\mathrm{cm^{-3}}$, 
respectively. 
We note that our ratio $f/i$ for O~{\sc vii} is somewhat lower than
that obtained by \citet{Brinkman2000} from LETGS spectra.  The HETGS and
LETGS observations were not simultaneous; however, based on evidence
from prior EUVE observations 
\citep{Dupree00}, we would be surprised if this difference
represented actual changes in the mean coronal plasma density.
Instead, we suggest that this results from different treatements
of the continuum plus background,
which particularly affects the strength of the intercombination line.





\section{Discussion}

These X-ray data confirm that Capella's corona contains plasma
at multiple temperatures in the accessible range from $\log T \sim
6.3$ to $7.2$,  and set stringent
constraints on the amount of plasma hotter than $\log T=7.2$ at the
time of this observation. These properties are generally consistent
with the results found with EUVE and ASCA \citep{BRICK00} and the line
strengths are close those seen 20 years earlier
by \citet{Vedder83}.

The preliminary results presented here have implications for the
structure of Capella's corona: they suggest that the characteristic
dimensions of the coronal loops at $T \sim 2 \times 10^6 \mathrm{K}$
are small compared to the stellar radius, $R_*$.  For simple
semi-circular loops of constant circular cross-section of radius $r$, we
use the measured density and $VEM$ for oxygen to estimate loop heights
$\leq0.02R_*\alpha_{0.1}^{-2/3}N_{100}^{-1/3}$, where $\alpha_{0.1}$ is
the ratio of $r$ to loop length in units of 0.1, and $N_{100}$ is
$1/100$ the number of loops. 
Detailed loop modeling of \citep{vdOord97} also required compact
structures, though variable cross-section loops were needed to
increase the proportion of hot to cool plasma.

\acknowledgements Work at MIT was supported by NASA through the HETG
contract NAS8-38249 and through Smithsonian Astrophysical Observatory
(SAO) contract SVI--61010 for the Chandra X-ray Center (CXC).  JJD and
NSB were supported by NASA NAS8-39083 to SAO for the CXC.  We thank
all our colleagues who helped develop the HETGS and all members of the
Chandra team.



\newcommand{\CapImg} { The HETGS spectrum is shown as an image in sky
  coordinates (top).  The zeroth order image is in the center (the
  vertical streak is caused by the CCD readout). The orientations of
  the MEG and HEG gratings are offset so the spectra form a shallow
  ``x'' on the ACIS array.  The sky image has been blurred for visual
  presentation, since the distribution is too narrow to see features.
  Below is a small portion of each spectrum (also smoothed) as
  labeled.
  \label{fig:image}
  }

\newcommand{\CapSpec}{
  The merged MEG $\pm1$ order spectrum is shown together with two
  insets: on the left, a region which compares the HEG (thin line) to
  MEG; the O~{\sc vii} He-like triplet is on the right.
  \label{fig:spectrum}
  }

\newcommand{\CapVEM}{
  The volume emission measure distribution based on analysis of   
  our HETGS spectra and from earlier work.  The symbols mark the $VEM$
  that would correspond to the case where all the observed emission
  for the line originates in an isothermal plasma with a temperature
  corresponding to the peak in the emissivity for that line.
  These points are shown for selected lines from
  Table~\ref{tab:linelist} spanning a range in
  formation temperature.  For a few of these, curved lines are the
  loci corresponding to the $VEM$ required to produce the observed
  flux from an isothermal plasma as a function of 
  plasma temperature.  The dashed line is an upper limit based on the
  observed counts in the vicinity of the Fe {\sc xxv} resonance line
  (1.85\AA).  The solid line is the emission measure distribution
  derived from ultraviolet and extreme ultraviolet emission lines by
  \citet{Brick96} plotted for $d\log T=0.1$.
  \label{fig:vemt}
  }





\clearpage

{\small
\begin{deluxetable}{rrrrr}
\tablewidth{0pt}
\tablecaption{Selected emission lines from MEG.\label{tab:linelist}}
\tablehead{
  \colhead{Line}&
  \colhead{$\lambda$\tablenotemark{a}}&
  \colhead{Flux\tablenotemark{b}}&
  \colhead{$C$\tablenotemark{c}}&
  \colhead{$T_\mathrm{m}$\tablenotemark{d}}
  }
\startdata
Fe \sc xxv&  1.85 &  $<4$&  $<7$   &  7.8\\        
S \sc xv&     5.040&   33&   110  &  7.2\\         
S \sc xv&     5.060&   16&   55   &  7.2\\        
S \sc xv&     5.100&   26&   85   &  7.2\\        
Si \sc xiii&  5.680&   25&   93   &  7.0\\        
Si \sc xiv&   6.180&   48&   472  &  7.2\\        
Si \sc xiii&  6.650&  182&   1228 &  7.0\\          
Si \sc xiii&  6.690&   47&   374  &  7.0\\        
Si \sc xiii&  6.740&  121&   834  &  7.0\\        
Al \sc xii&   7.750&   16&   204  &  6.9\\           
Mg \sc xii&   8.419&  152&   1947 &  7.0\\          
Mg \sc xi&    9.170&  348&   2818 &  6.8\\          
Mg \sc xi&    9.230&   63&   618  &  6.8\\         
Mg \sc xi&    9.310&  190&   1425 &  6.8\\          
Ne \sc x&     10.240&   92&  740  &  6.8\\         
Ni \sc xxii&  10.791&   62&  426  &  7.0\\        
Ne \sc x\tablenotemark{e}&     12.132&  929&  4171 &  6.8\\          
Fe \sc xvii\tablenotemark{e}&12.134& \nodata& \nodata&  6.8\\ 
Fe \sc xix\tablenotemark{e}&   13.515&  530&   1587 &  6.9\\            
Fe \sc xix\tablenotemark{e}&   13.524& \nodata& \nodata&  6.9\\            
Fe \sc xvii&  15.013& 3043&   7476 &  6.7\\            
Fe \sc xvii&  15.272& 1119&   2919 &  6.7\\                      
Fe \sc xviii& 15.641&  410&   938  &  6.8\\         
O \sc viii& 16.003&  898&   1885 &  6.5\\          
Fe \sc xvii&  16.796& 2004&   3669 &  6.7\\          
Fe \sc xvii&  17.071& 2641&   4554 &  6.7\\          
Fe \sc xvii&  17.119& 2443&   4191 &  6.7\\          
O \sc viii&   18.967& 2634&   2810 &  6.5\\          
O \sc vii&    21.600&  967&   396  &  6.3\\         
O \sc vii&    21.800&  255&   102  &  6.3\\        
O \sc vii&    22.100&  736&   249  &  6.3\\        
N \sc vii&    24.779&  549&   327  &  6.3\\        
\enddata
\tablenotetext{a}{Theoretical wavelengths of identification, in \AA.}
\tablenotetext{b}{Observed flux is
  $10^{-6}\times$ the tabulated Flux in $[\mathrm{phot\,cm^{-2}\,s^{-1}}]$.
  The systematic uncertainty is less than 10\% except in the range
  from 6--12\AA\  where it is up to   20\%.}  
\tablenotetext{c}{$C$ is the integrated line counts in 89 ks.} 
\tablenotetext{d}{$T_\mathrm{m}$ is the $\log$ temperature (in K) of
  maximum emissivity.}
\tablenotetext{e}{Blend; entry for sum of two components}
\end{deluxetable}
}


\onecolumn
\begin{figure}
 \figurenum{1}
\epsscale{1.0}
\plotone{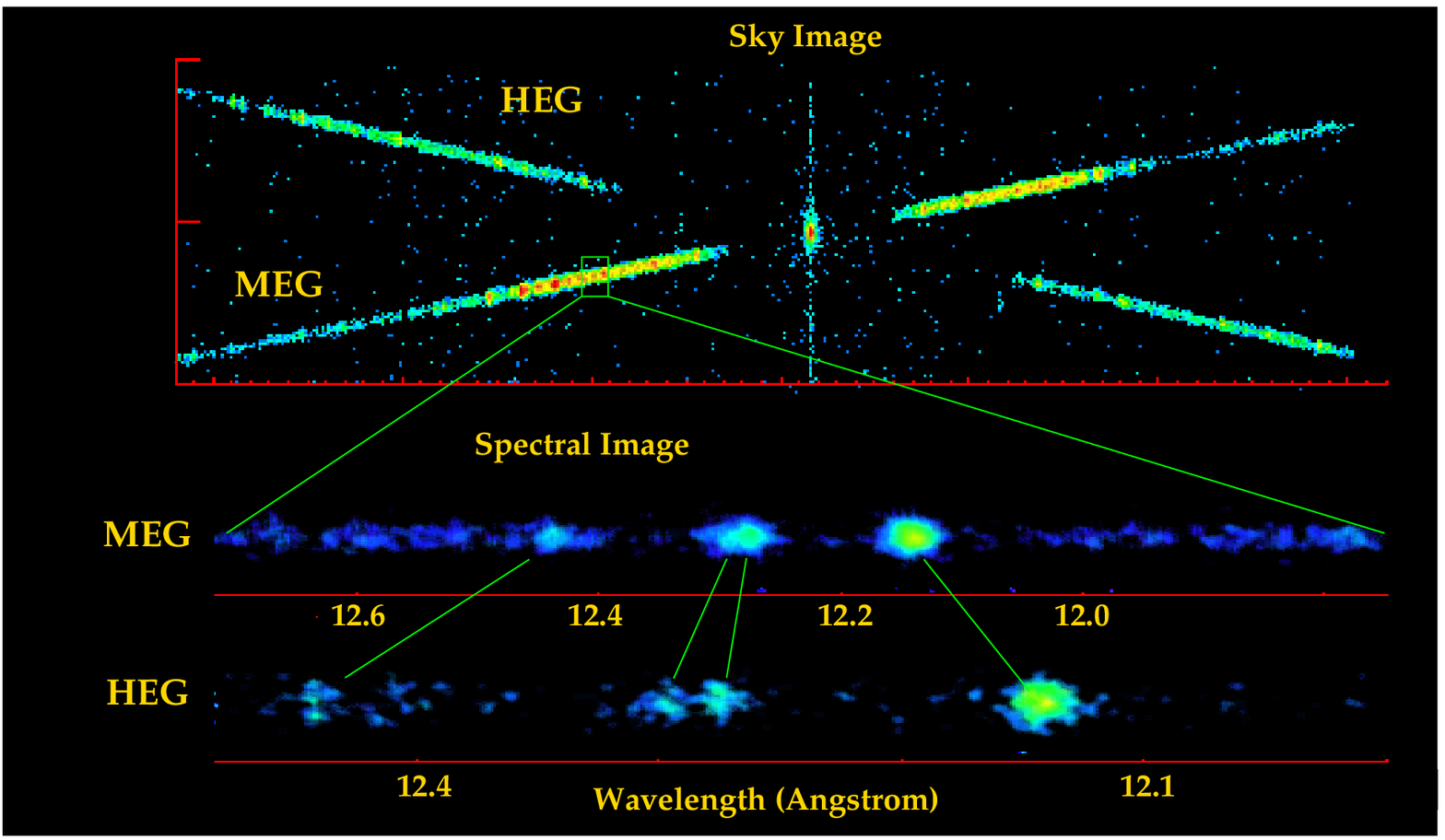}
\caption{\CapImg}
\label{fig:image}
\end{figure}
\begin{center}
\begin{figure}[h]
\figurenum{2}
 \epsscale{1.0}
\plotone{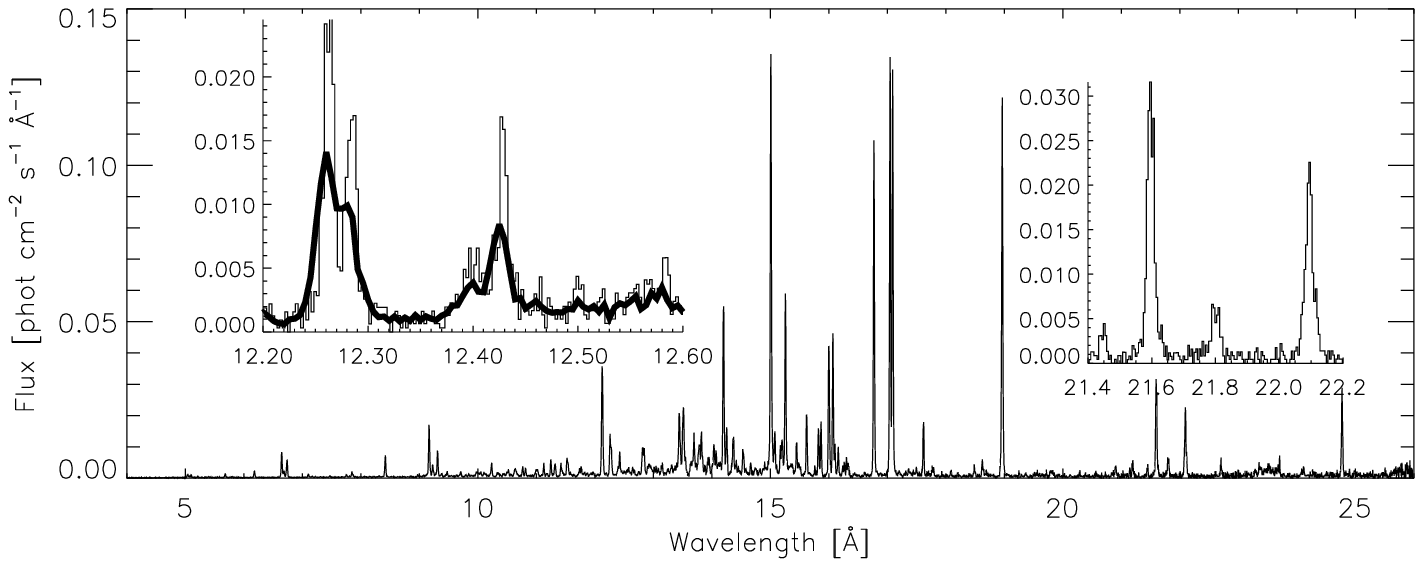}
\caption{\CapSpec}
\label{fig:spectrum}
\end{figure}
\end{center}
\begin{center} 
\begin{figure}[h]
\figurenum{3} 
\epsscale{1.0} 
\plotone{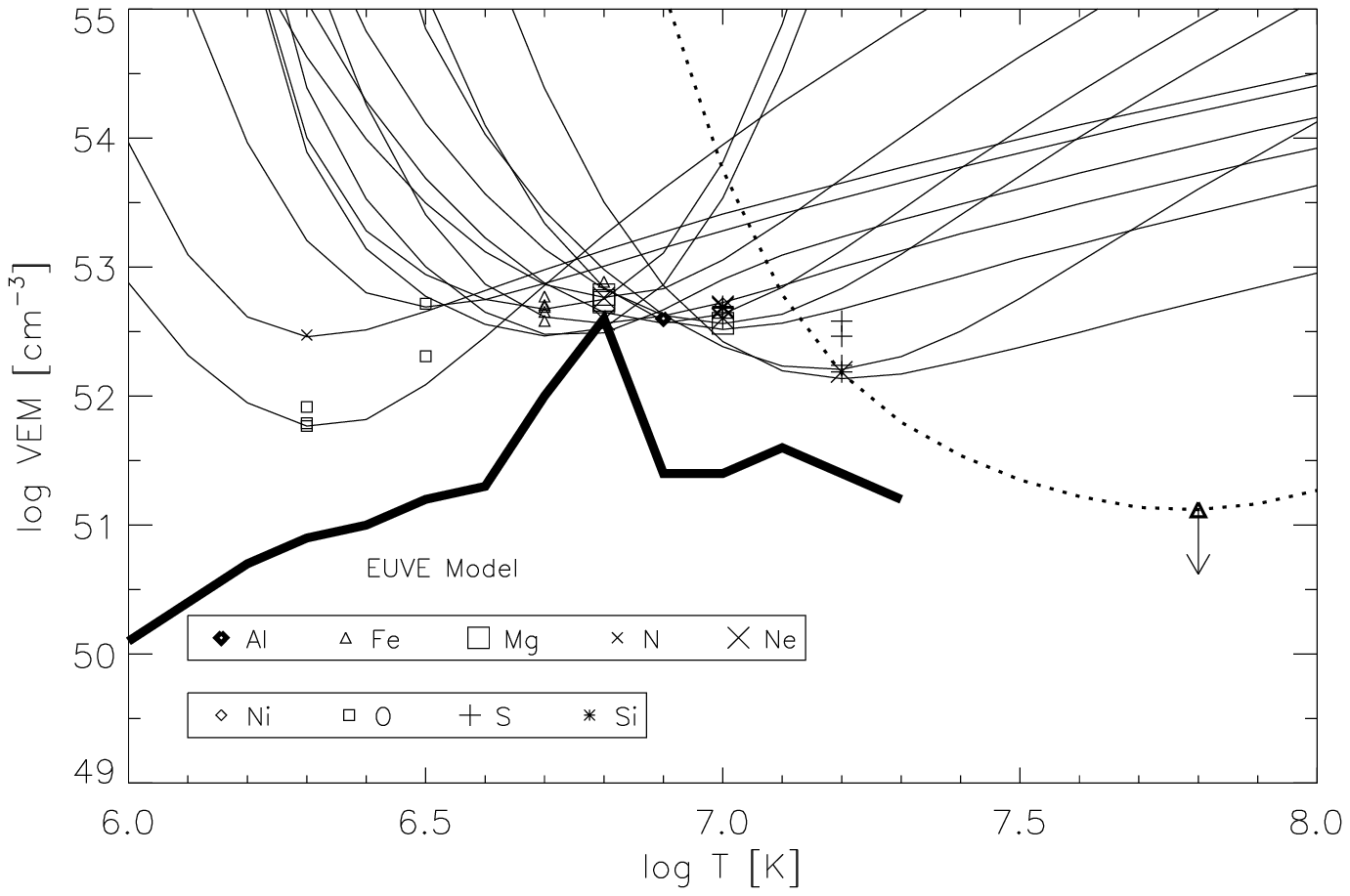}
\caption{\CapVEM}
\label{fig:vemt}
\end{figure}
\end{center}


\frenchspacing
\begin{thebibliography}{}
%
%
\bibitem[Anders and Grevesse(1989)]{Anders89}
  Anders, E., \& Grevesse, N., 1989,
  Geochimica et Cosmochimica Acta 53, 197.
%
\bibitem[Brickhouse(1996)]{Brick96} %
  Brickhouse, N. S. 1996, in Proc. IAU Colloq. 152,
  {\it Astrophysics in the Extreme Ultraviolet\/},
  ed. S. Bowyer \& R. F. Malina, (Dordrecht: Kluwer), 105.
%
\bibitem[Brickhouse et al.(2000)]{BRICK00} %
Brickhouse, N.~S., Dupree, A.~K., Edgar, R.~J., Liedahl, D.~A., Drake,
S.~A., White, N.~E., \& Singh, K.~P. 2000, \apj, 387.
%
\bibitem[Brinkman, et al.(2000)]{Brinkman2000} 
  Brinkman, A.C., et al., 2000, \apjl, (submitted)
%
\bibitem[Brown, et al.(1998)]{Brown98} 
  Brown, G.V., Beiersdorfer, P., Liedahl, D.A., et al., 1998, \apj,
  502, 1015.
%
\bibitem[Canizares, et al.(2000)]{Canizares00}
        Canizares, C.R. et al. 2000 in preparation
%
%
\bibitem[Dempsey et al.(1993)]{Dempsey93b}
  Dempsey, R.C., Linsky, J.L., Schmitt, J.H.M.M., and Fleming,T.A.,1993,
  \apj,413,333
%
%
\bibitem[Dupree et al.(1993)]{Dupree93}
  Dupree, A.K., Brickhouse, N.S., Doschek, G.A., Green, J.C., and Raymond,
  J.C., 1993, \apj, 418, L41
%
\bibitem[Dupree et al.(2000)]{Dupree00}
  Dupree, A.K., Brickhouse, N.S., and Sanz-Forcada, J., 
  2000, in preparation.
%
\bibitem[Favata et al.(1997)]{Favata97}
Favata, F., Mewe, R., Brickhouse, N. S., Pallavicini, R.,
Micela, G., \& Dupree, A. K. 1997, \aap, 324, L37
%
\bibitem[Gabriel(1972)]{Gabriel72}
Gabriel, A.~H. 1972, \mnras, 160, 99
%
\bibitem[Gabriel and Jordan(1969)]{Gabriel69}
Gabriel, A.~H., \& Jordan, C. 1969, \mnras, 145, 241
%
\bibitem[Griffiths and Jordan(1998)]{GRIFF98} %
  Griffiths, N.W., Jordan, C., 1998 \apj, 497, 883
%
%
\bibitem[Holt et al.(1979)]{Holt79}
Holt, S. S., White, N. E., Becker, R. H., Boldt, E. A.,
Mushotzky, R. F., Serlemitsos, P. J., \& Smith, B. W. 1979, \apj,
234, L65
%
\bibitem[Hummel et al.(1994)]{Hummel94}
Hummel, C. A., Armstrong, J. T., Quirrenbach, A., Buscher,
D. F., Mozurkewich, D., \& Elias~II, N. M. 1994, \aj, 107, 1859
%
%
%
\bibitem[Lemen et al.(1989)]{Lemen89}
Lemen, J. R., Mewe, R., Schrijver, C. J, \& Fludra, A. 1989,
\apj, 341, 474
%
\bibitem[Linsky et al.(1998)]{Linsky98}
  Linsky, J.L., Wood, B.E., Brown, A., \& Osten, R.A., 1998,
  \apj, 492, 767
%
\bibitem[Markert et al.(1994)]{Markert94}
Markert, T.H., Canizares, C.R., Dewey, D.,
McGuirk, M., Pak, C.S., \& Schattenburg, M.L., 1994,
Proc. SPIE, 2280, 168

%
\bibitem[Mewe et al.(1982)]{Mewe82}
  Mewe, R. et al. 1982, \apj, 260, 233
%
\bibitem[Pradhan \& Shull (1981)]{Pradhan81}
  Pradhan, A. K. and Shull, J.M 1981, \apj, 249, 821
%
%
%
\bibitem[Saba et al.(1999)]{Saba99}
  Saba, J.L.R., Schmelz, J.T., Bhatia, A.K., and Strong, K.T., \apj,
  510, 1064.
%
\bibitem[Schmitt, et al.(1990)]{Schmitt90} 
  Schmitt, J.H.M.M., Collura, A., Sciortino, S., Vaiana, G.S.,
  Harnden, F.R., Jr., and Rosner, R., 1990, \apj, 365, 307.
%
\bibitem[Schrijver et al.(1995)]{Schrijver95}
  Schrijver, C.J., Mewe, R., van den Oord, G.H.J., and Kaastra, J.S. 
  1995, \aap, 302, 438.
%
\bibitem[Smith et al.(1998)]{Smith98}
  Smith, R.~K., Brickhouse, N.~S., Raymond, J.~C., \& Liedahl, D.~A. 1998,
  in Proceedings of the First XMM Workshop on ``Science with XMM'',
  ed.\ M.~Dahlem (Noordwijk, The Netherlands)
%
\bibitem[Strassmeier, et al.(1993)]{CABS93}
  Strassmeier, K.G., Hall, D.S., Fekel, F.C., and Scheck, M., 1993, 
  \aaps, 100, 173 
%
\bibitem[Swank et al.(1981)]{Swank81a} 
  Swank et al.,  1981, \apj, 246,  214 
%
\bibitem[van den Oord et al.(1997)]{vdOord97} 
van den Oord, G.H.J., Schrijver, C.J., Camphens, M., Mewe, R., \&
Kaastra, J.S.,
\aap, 326, 1090
%
%
\bibitem[Vedder and Canizares(1983)]{Vedder83}  %
  Vedder, P.W., Canizares, C.R., 1983, \apj, 270, 666
%
%
%
%
\end{thebibliography}
\end{document}